\documentclass[prb,showpacs,preprint]{revtex4}
\usepackage {graphicx}

\begin{document}

\title{Mapping of surface local elastic constant using Atomic Force Acoustic Microscopy}

\author{S. Banerjee\footnote{Corresponding author and on Deputation from Surface Physics Division, Saha Insitute of Nuclear Physics, 1/AF Bidhannagar, Kolkata, Email:sangam@hp1.saha.ernet.in}, N. Gayathri, S.~R.~Shannigrahi$^\#$, S. Dash, A. K. Tyagi and B. Raj}

\address{Materials Science Division, Indira Gandhi Centre for Atomic Research, 
Kalpakkam – 603102, T.N, India}

\address{$^\#$Institute of Materials Research and Engineering, 3 Research Link, 117 602 Singapore}

\begin{abstract}
We report a systematic study to determine local elastic properties of surfaces combining atomic force microscope (AFM) with acoustic waves which is known as atomic force acoustic microscopy - AFAM. We describe the methodology of AFAM in detail and interpret the measurement using simple arguments and other complementary measurements using AFM. We have used a few selected samples to elucidate the capability of the AFAM technique to map the local elastic constant of the sample surface. Using the force-distance measurement and the change in the frequency of resonance peak at various regions we quantitatively determine the change in the local stiffness (elastic constant) of the sample surface. We have also shown that AFAM technique can be used to get a better surface image contrast where contact mode AFM fails. 
\end{abstract}
\maketitle

\section{Introduction}
Atomic force microscopy (AFM) is now one of the major microscopic techniques used to characterize nanoscale structures on surfaces and hence has become an indispensible characterzation tool in the field of nanoscience and nanotechnology, being routinely used in different fields. In the AFM technique a flexible micro-fabricated elastic cantilever carrying a sharp tip is used which acts as a {\it force~sensing} probe. These forces can be electric or magnetic in origin. Hence, AFM can be used as a non-destructive tool to characterize surfaces of its rms roughness, frictional coefficient, nano-hardness, electrical and magnetic field mapping of surface and many other surface properties by using appropriate cantilevers. These cantilevers are generally elastic beam of various shapes and by means of classical elastic theory one can calculate the properties of its bending, resonance frequencies, flexural and torsional modes of vibration. By studying the elastic properties of the cantilever when the tip is in contact with the sample surface one can determine the surface characteristics of the sample. However, the AFM technique is still being improvised to influence the cantilever by various means for getting further information of the samples. Atomic force acoustic microscopy (AFAM) is one such improvised AFM technique which has recently been developed \cite{Cretin,Yamanaka,Rabeapl94,Rabersi} and is still being improvised \cite{kolosovprl,Hurley1,Hurley2,Dinelli}. We will show in the present study that AFAM is an emerging novel technique to map local elastic properties of sample surfaces. In the present study we will also use force vs. distance measurement and resonance curves with the tip in contact with the sample surface to estimate quantitatively relative change in local elastic constants on the surface of the sample.

\section{Principle of AFAM}

AFAM technique has been described earlier \cite{Rabeus,Rabess} and versions with slight variations were also proposed by others \cite{Yamanakaapl,kolosovprl}. Here we will describe the AFAM measurement technique in detail and provide a simple explanation and measurements for interpreting the results quantitatively. The experimental setup is shown schematically in Fig.1. In this technique the sample is bonded by any ultrasonic bonding material to a longitudinal ultrasonic piezoelectric transducer. In the AFAM, first, the tip is approached towards the sample with the feedback ON to make contact with the sample surface. This is done to assure that the interaction between the tip and the sample surface is around a constant value through out the experimental scan as is done in the conventional AFM to get topography image in contact mode. One can increase or decrease the tip-sample interaction value by changing the amount of deflection of the cantilever which determines the tip-sample interaction. The amount of deflection is monitored by the DC signal of the photodiode. In other words, the amount of deflection of the laser beam from the centre of the photodiode determines the applied force on the sample (tip-sample interaction) (see Fig. 1). In AFAM mode one has to optimize this force (interaction strength) such that surface modification does not take place in case of ultra thin fragile films. We have earlier shown that the surface of the sample can be modified deliberately to make self assembled nano-patterns and quantum dots \cite{jpd1,epjap,jpd2} with an AFM tip. After the tip is made to contact with the sample the ultrasonic transducer bonded to the sample is excited (typical resonance frequency of the first harmonic of the ultrasonic transducer is around 1 MHz). The excitation voltage should be kept low such that the resonance curve is undistorted and is typically kept less than 1V. The tip-sample-transducer coupled system will have a resonance frequency which will depend on the effective elastic constant E* between the tip and the sample surface (see Fig.1), which may vary as a function of position of the tip on the sample surface. The amplitude of vibration of the cantilever induced by the longitudinal vibration of the ultrasonic transducer via the sample is obtained by monitoring the amplitude of the AC component of the photodiode signal using a lock-in amplifier as shown in Fig. 1. As one sweeps the excitation frequency of the ultrasonic transducer around its resonance peak one obtains a resonance curve of the whole coupled (tip-sample-transducer) system. Depending on whether the effective elastic constant between the tip and the sample surface is higher (stiff) or lower (soft) the peak of the resonance curve will shift towards a higher or lower frequency respectively. This is schematically shown in Fig. 2. Now, let the excitation frequency of the transducer which is also fed as the reference frequency of the lock-in amplifier be set slightly above the peak value (i.e., on the right hand side of the resonance peak) as shown in Fig. 2. Now, if the tip is in contact with a higher elastic constant region on the surface of the sample this will cause the amplitude of oscillation of the cantilever to increase (marked by arrow 1) and if the tip is in contact with a lower elastic constant region, then the amplitude of oscillation of the cantilever will decrease (marked by arrow 2). Thus increase in amplitude of oscillation of the cantilever corresponds to a higher elastic constant region and decrease in amplitude of oscillation corresponds to a lower elastic constant region. If this is true and if we now choose the transducer excitation frequency and hence the reference frequency for the lock-in amplifier to the left of the resonance peak then the situation should be reversed i.e., the region corresponding to higher elastic constant will cause a decrease in the amplitude of oscillation of the cantilever because now the resonance curve will move towards the right i.e., towards higher frequency (marked by arrow 3) and the region having lower elastic constant will exhibit increase in amplitude of oscillation since the resonance curve moves left i.e., toward the lower frequency (marked by arrow 4) as shown schematically in Fig. 2. Thus in the former case (i.e., when the frequency is chosen above the peak value) if the higher amplitude of oscillation is designated to be a brighter shade and lower amplitude to be a darker shade then one can map the relative change in local elastic constant on the surface of the sample where the bright region signifies stiff region and the dark region signifies soft region. In the later case (i.e., when the frequency is chosen below the peak value) the scenario will be reversed. We will show in this investigation that this infact is observed and from this we can map the distribution of higher and lower elastic constant regions on the surface of the sample. We will also carry out force vs. distance curve for soft (low elastic constant region) and stiff (high elastic constant region) and show that the resonance peaks shifts for these two regions to confirm our model to explain the AFAM image.

	To add few more points, we can also obtain the ratio of the elastic constants for corresponding two regions by calculating the average stiffness constant $k^*$ of the tip-sample system using Hertz model \cite{Israel}

\begin{equation}
k^* = \sqrt[3]{6{{E^*}^2}RF_N}
\end{equation}

\noindent where $R$ is the radius of curvature of the tip and $F_N$ is the normal force acting on the sensor tip which can be obtained from force-distance curve and $E^*$ can be expressed as

\begin{equation}
\frac{1}{E^*} = \frac{{1-\nu_t}^2}{E_t}+\frac{{1-\nu_s}^2}{E_s}
\end{equation}  								

\noindent where $E_t$, $\nu_t$, $E_s$ and $\nu_s$ are the Young's modulus the Poisson's ratio of the tip and the sample surface respectively. We would like to point out that in Hertz model the tip and sample surface are considered to be rigid bodies and we have shown recently that in certain cases the tip-sample interaction is generally described by JKR model where the tip and the sample surface are considered to be deformable bodies and in this case both the sample surface and the tip gets modified \cite{jpd2}. In Hertz model, the ratio of the elastic constant among two regions can now be expressed as 

\begin{equation}
\frac{{E_1}^*}{{E_2}^*} = [\frac{{k_1}^*}{{k_2}^*}]^{3/2}
\end{equation}  								
 
\noindent where the subscript 1 and 2 denotes region 1 and 2 respectively.

In this study we will also quantitatively evaluate the ratio of elastic constants as mentioned above using the force vs. distance curve for soft and stiff regions and also using the resonance peaks at those regions to demonstrate that indeed we map stiff and soft region using AFAM technique.

\section{Samples studied}

For the present investigation we have selected three samples (1) polished commercial piezoelectric PZT, Pb(Zr,Ti)O$_3$ ceramic, (2) Thin film of PZT deposited by sol-gel technique and (3) Thin film of Au deposited on Si. The first sample is used to study and demonstrate the AFAM technique as described above to obtain the contrast in elastic constant and compare it with the topography image. The second sample is used to show how AFAM can be used to get better contrast than the topography image using AFAM technique and the third sample is used to determine quantitatively some elastic constant parameters as described above. The experiments were carried out using a commercial (NT-MDT, Russia) SPM setup consisting of waveform generator and built-in lock-in amplifier.

\section{Results and discussion}

In Fig. 3 we show topography and AFAM images of the PZT sample. In Fig. 3(a) and 3(b) we show topography image using contact mode and AFAM image taken with ultrasonic frequency chosen $above$ the contact resonance peak frequency (see fig. 2) and using this frequency as the reference to the lock-in amplifier. The topography does not show much features and the mean roughness of the sample is ~ $<$ 10 \AA. On the other hand in the case of AFAM image we observe distinct bright and dark patterns where the bright pattern as explained above can be attributed to stiff regions and the dark patterns to the softer regions on the surface of the sample. As discussed above (in section II), if we now select the ultrasonic frequency $lower$ than the contact resonance peak frequency and use this frequency as the reference to the lock-in amplifier we then observe inversion of the dark and bright pattern as shown in Fig. 3(d). The topography images Fig. 3(a) and 3(c) which is just the DC feedback signal does not show very clear features and it only represents the height distribution of the surface. Whereas, the bright and dark pattern observed by monitoring the amplitude of the AC component of the cantilever can only be explained by invoking the idea of distribution of soft and stiff region. The dark and bright regions are the individual ferroelectric domains within the polycrystalline multi-domain sample. Since the polarized domains are oriented in different axis and directions the elastic constant along that axis and direction are also different and hence we see such distribution of ferroelectric domains on the sample surface. This type of domain structure for PZT material has also been observed earlier by Rabe et al. \cite{Rabeus}, but here we explain the scenario using the idea described in Fig. 2. This describes the phenomenon  more pictorially and presents a more easier understanding. More conclusive evidence that AFAM really maps the soft and the stiff regions will be demonstrated using our third sample.

We would like to show now that this technique can also be very useful to achieve better contrast of the surface images which other wise is difficult to obtain using contact mode AFM. We now carried out the AFAM technique to characterize thin film of PZT grown by sol-gel technique. In Fig. 4(a) we show the topography image and in Fig. 4(b) we show the AFAM image with ultrasonic frequency above the resonance peak value. We can clearly see the differences in the contrast between the topography image and the AFAM image. We see from both the images that the PZT film grown by sol-gel technique shows mound formation. We observe that, when the operating frequency is above the peak value the peripheral region of the mounds appears to be brighter than the regions within mounds which appears to be darker. If we now select the operating frequency lower than the peak value as was done in our earlier case then we observe the reversal of the bright and dark region i.e., we observe a contrast inversion as shown in Fig. 4(d). Thus we can conclusively say that the peripheral regions of the mound is stiffer than within the mounds. This is an interesting observation indicating that the mound can be easily compressed at its center than at the periphery.

To quantify the elasticity we have carried out AFAM measurements on a thin film of Au coated on Si substrate. We have selected a region where we have deliberately scratched the surface. The topography is shown in Fig. 5(a) and the AFAM images are shown in Fig. 5(b) and (c) with operating frequency $above$ the contact resonance peak value and with operating frequency $below$ the contact resonance peak value respectively. As in earlier case we observe contrast inversion. We can see that material dugged out by scratching which lies along the ridge (we call it a debris region) is softer because we see it as dark region in Fig. 5(b). On the other hand the ridge (i.e., the middle portion of the scratch) from where the material has been removed appears to be stiff because it appears bright. Now if we change the operating frequency below the resonance peak we then observe opposite contrast in the AFAM image as shown in Fig. 5(c). To make it more certain about our conclusions we have measured the force vs. distance curve on the unscratched region marked in Fig 5(a) by cross and on the peeled debris region marked by circle in the same figure. The force vs. distance curves taken on both the regions are shown in fig. 6(a). Around each region we have taken five curves and Fig. 6(a) shows their average. For obtaining a force vs. distance curve one generally retreats the z-piezo either attached to the cantilever or the sample and then push forward beyond the contact point of the tip-sample upto the value set by the operator. One can measure force vs. distance curve with different initial condition of the z-piezo. In Fig. 6(a) we show force vs. distance curve for two different initial condition (sample-cantilever holder seperation). If the cantilever holder-sample seperation is small then the tip will contact the sample earlier as shown in Fig. 6(a) marked by arrow 1 for curves B and if the seperation is larger it will contact later as marked by arrow 2 for curves A. We observe that the slope of the force vs. distance curve for the debris region is lower (solid lines) than the slope of curve obtained on the unscratched region (i.e., the plain region, dot-dash lines). This indicates that the debris region is softer than the plain region. This observation is similar to that we have observed from the AFAM image. The y-axis of the curve in Fig. 6(a) is plotted as normalised force. The normal force acting on the tip by the sample is the same as the applied force (load) by the tip on the sample. The bending of the cantilever is proportional to this force. Considering static conditions and at equilibrium the sample-cantilever force is just simply the restoring force of the cantilever and hence can be expressed as (Hooke's law)

\begin{equation}
F_N(z) = k^* \Delta z
\end{equation}

\noindent where $k^*$ is the effective stiffness constant. If $F_1$ and $F_2$ are the forces in region 1 and region 2, then they can be expressed as  $F_{N,1} = k_1 ^* \Delta z_1$ and $F_{N,2} = k_2 ^* \Delta z_2$. If we now obtain $\Delta z_1$ and $\Delta z_2$ for $F_{N,1} = F_{N,2}$ we can express the ratio of the effective stiffness constants as 

\begin{equation}
\frac{k_1 ^*}{k_2 ^*} = \frac{\Delta z_2}{\Delta z_1}
\end{equation}

\noindent From Fig. 6(a) we get $k_1 ^*/k_2 ^*$ = 1.030 for curve A and 1.027 for curve B. By taking their average, the ratio of the elastic constants becomes 1.043 using eqn.(3). This indicates that there is $\sim$ 4.3\% change in the local elastic constant among these two regions.

We show the resonance curve taken on the plain region and on the debris region in Fig. 6(b). We observe for the debris region the resonance peak is lower and broader than in the plain region indicating that the debris is softer than the plain region. We can also obtain the ratio of the elastic constants from the peak positions of the curves. If we consider the point-mass model (a point-mass m connected to the cantilever with spring constant $k_c$ as shown schematically in Fig. 1), the free resonance $f_0$ = $\sqrt{\frac{k_c}{m}}$. When the tip is in contact with the sample, the tip-sample interaction can be represented by the stiffness constant $k^*$, then, the coupled cantilever-tip-sample stiffness constant is $k^*+k_c$ using the model shown schematically in Fig 1. The resonance frequency of this coupled system is now $f_{res} = \sqrt{\frac{k^*+k_c}{m}}$. The effective $k^*$ can be found from

\begin{equation}
\frac{f_{res}}{f_0} = \sqrt{\frac{k^*+k_c}{k_c}}
\end{equation}

\noindent Hence the ratio of the effective stiffness constants for regions 1 and 2 can be expressed as 

\begin{equation}
\frac{k_1 ^*}{k_2 ^*} = \frac{f_{res,1} ^2 -f_0 ^2}{f_{res,2} ^2 -f_0 ^2}
\end{equation}

\noindent From Fig. 6(b) we get $f_{res,1}$ = 1617 kHz and $f_{res,2}$ = 1595 kHz and with $f_0$ = 314.6 kHz, we get the ratio of the effective stiffness constants to be 1.032 and hence the ratio of the elastic constants of the two regions is 1.044. This indicates that there is $\sim$ 4.4\% change in the local elastic constant among these two regions which is in very good agreement with that obtained using the force-distance curves discussed above.

\section{Conclusion}

We have shown that AFAM technique can be used as a very powerful tool to map the local elastic constant of the sample surface within nano-scale order. The same technique can be used to enhance the contrast of the image of the sample surface where conventional contact mode AFM fails. To compare quantitatively the local elastic constant in different regions one can perform static force-distance measurement or perform dynamic measurements using ultrasonic excitation and measure the shift in the resonance curve. We have thus shown that the novel AFAM and AFM technique provides a unique opportunity to study local elastic properties of nanostructures which further determines its physical properties for design application purpose.

\newpage

\begin{figure}
\includegraphics*[angle=270,width=18cm]{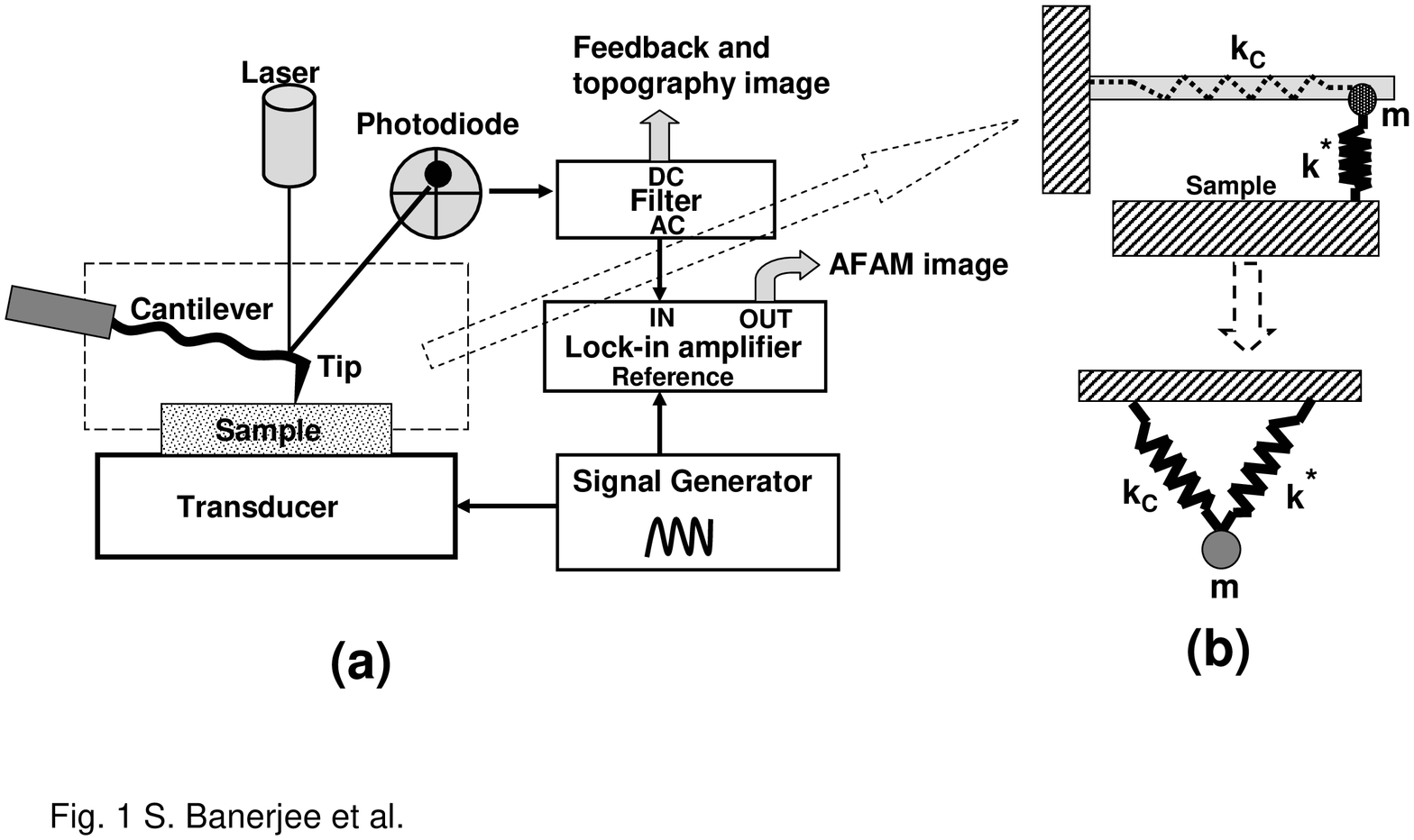}
\caption{(a) The schematic diagram of the experimental setup of Atomic Force Acoustic Microscopy (AFAM). The sample is bonded to an ultrasonic transducer and the transducer is excited by a waveform generator. The tip is in contact with the sample and the DC component of the photodiode is used for feedback and topography image and the AC component is used for AFAM image. (b) Schematic diagram describing cantilever-tip-surface coupled system, where $k_c$ is the stiffness constant of the cantilever and tip-surface interaction represented by a spring having effective stiffness constant $k^*$ and m is the mass of the tip considering point mass model.}
\end{figure}

\begin{figure}
\includegraphics*[angle=270,width=18cm]{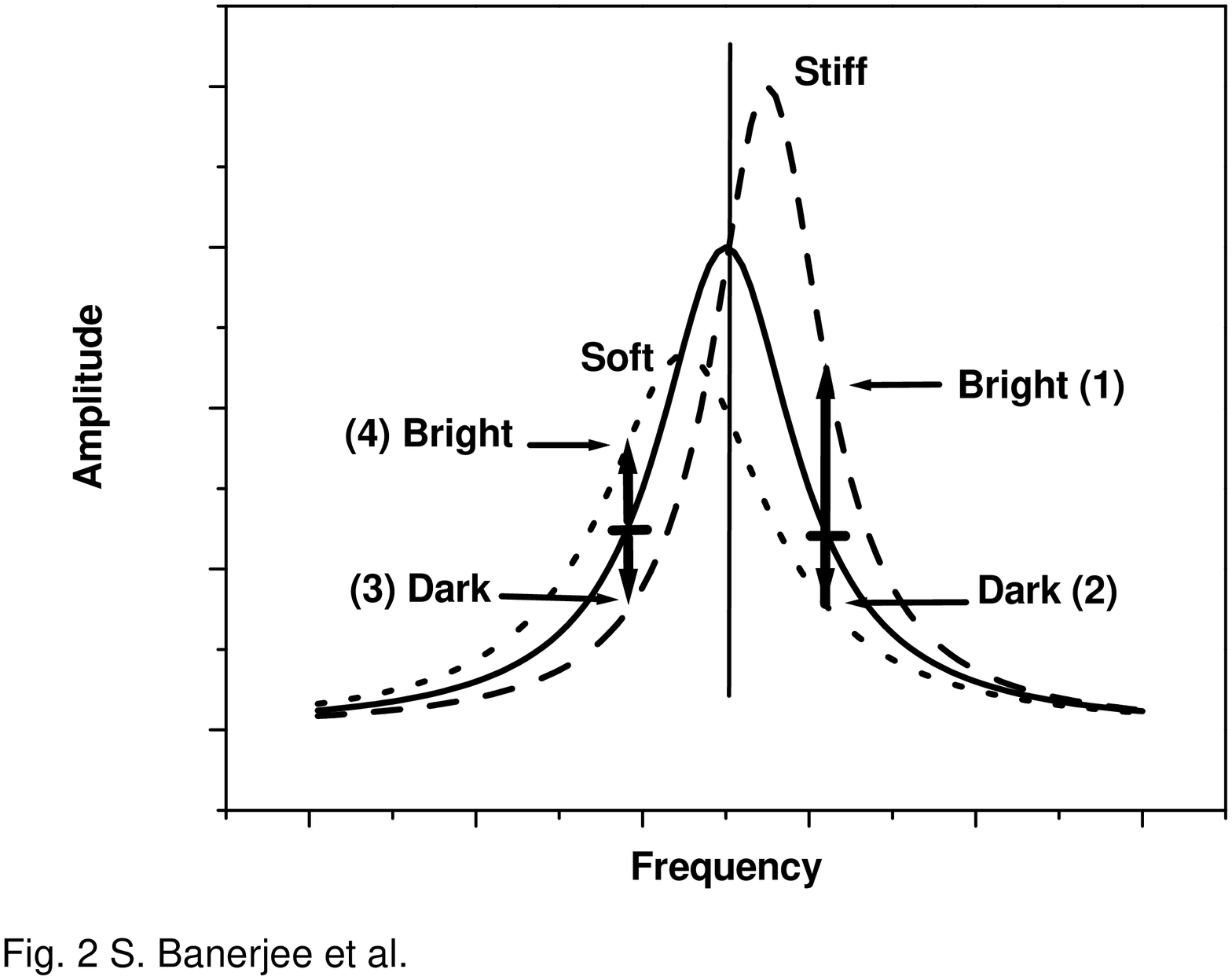}
\caption{Amplitude of oscillation of the cantilever as a function of excited frequency of the ultrasonic transducer while the tip is in contact with the sample. We show schematically the resonance curve and shift of the resonance curve for higher and lower effective elastic constant regions (marked as soft and stiff). Increase of oscillation amplitude corresponds to brighter image and decrease of the amplitude corresponds to darker image if the operating frequency is chosen above the resonance peak value and if chosen below then we observe contrast inversion. Note: As the local elastic constant increases the peak value of the amplitude and frequency increases and the FWHM decreases.}
\end{figure}

\begin{figure}
\includegraphics*[width=12cm]{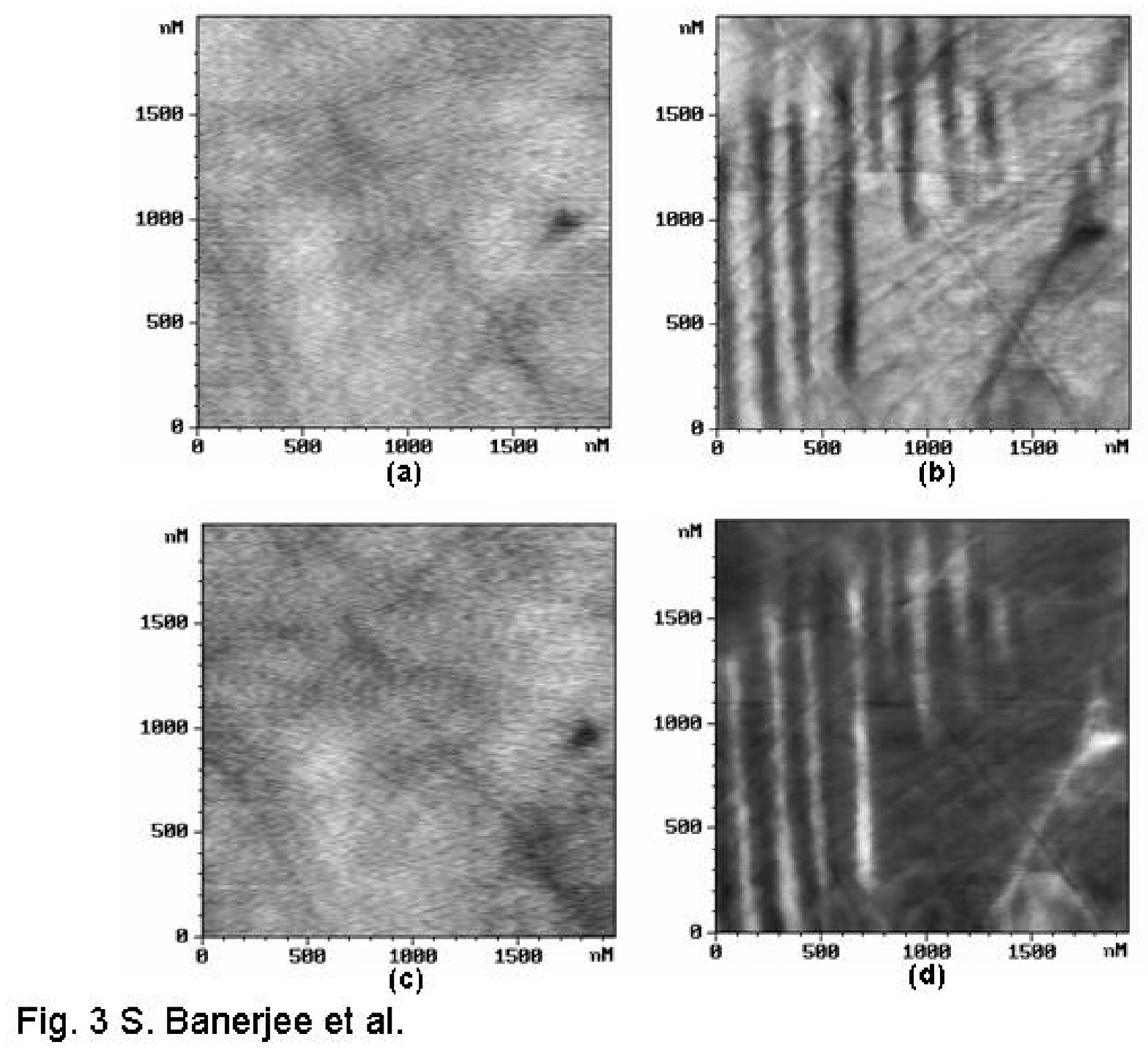}
\caption{We show topography and AFAM images of polished bulk sintered polycrystalline PZT sample. (a) and (b) are the topography and AFAM measurement carried out above the resonance curve as shown in Fig. 2, similarly (c) and (d) are the topography and AFAM measurement carried out below the resonance curve. The bright and dark patterns are the ferroelectric single domains (see text for details). We observe contrast inversion in (b) and (d)}
\end{figure}

\begin{figure}
\includegraphics*[width=12cm]{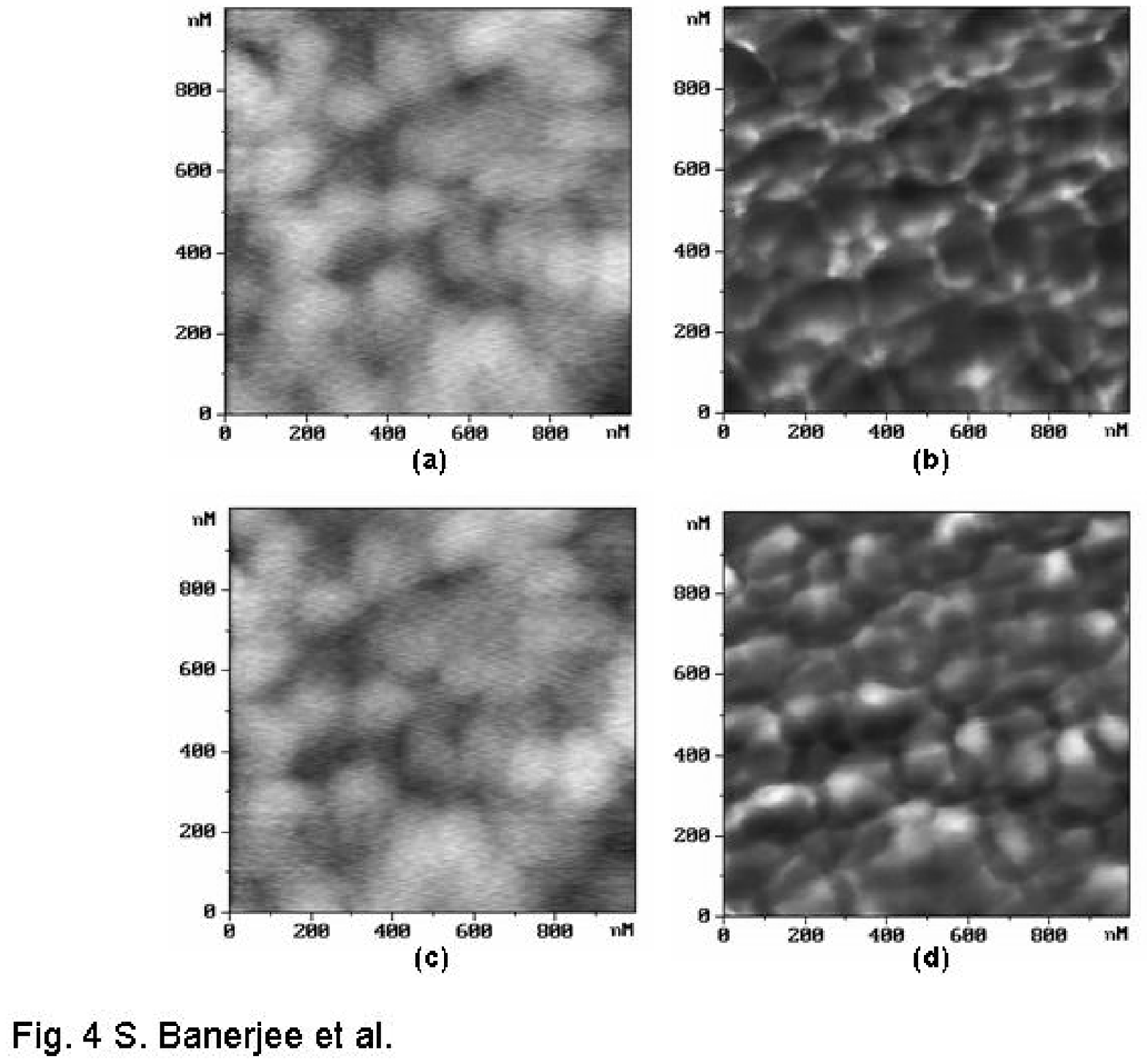}
\caption{We show topography and AFAM images of thin film of PZT sample grown by sol-gel technique. (a) and (b) are the topography and AFAM measurement carried out above the resonance curve, similarly (c) and (d) are the topography and AFAM measurement carried out below the resonance curve. In image (b) the bright regions corresponds to stiff regions and dark regions as soft regions. In image (d) the dark region corresponds to stiff region and the bright region corresponds to soft region. We observe contrast inversion in (b) and (d). The peripheral region of the mound is found to be stiffer than the region within the mounds.}
\end{figure}

\begin{figure}
\includegraphics*[width=12cm]{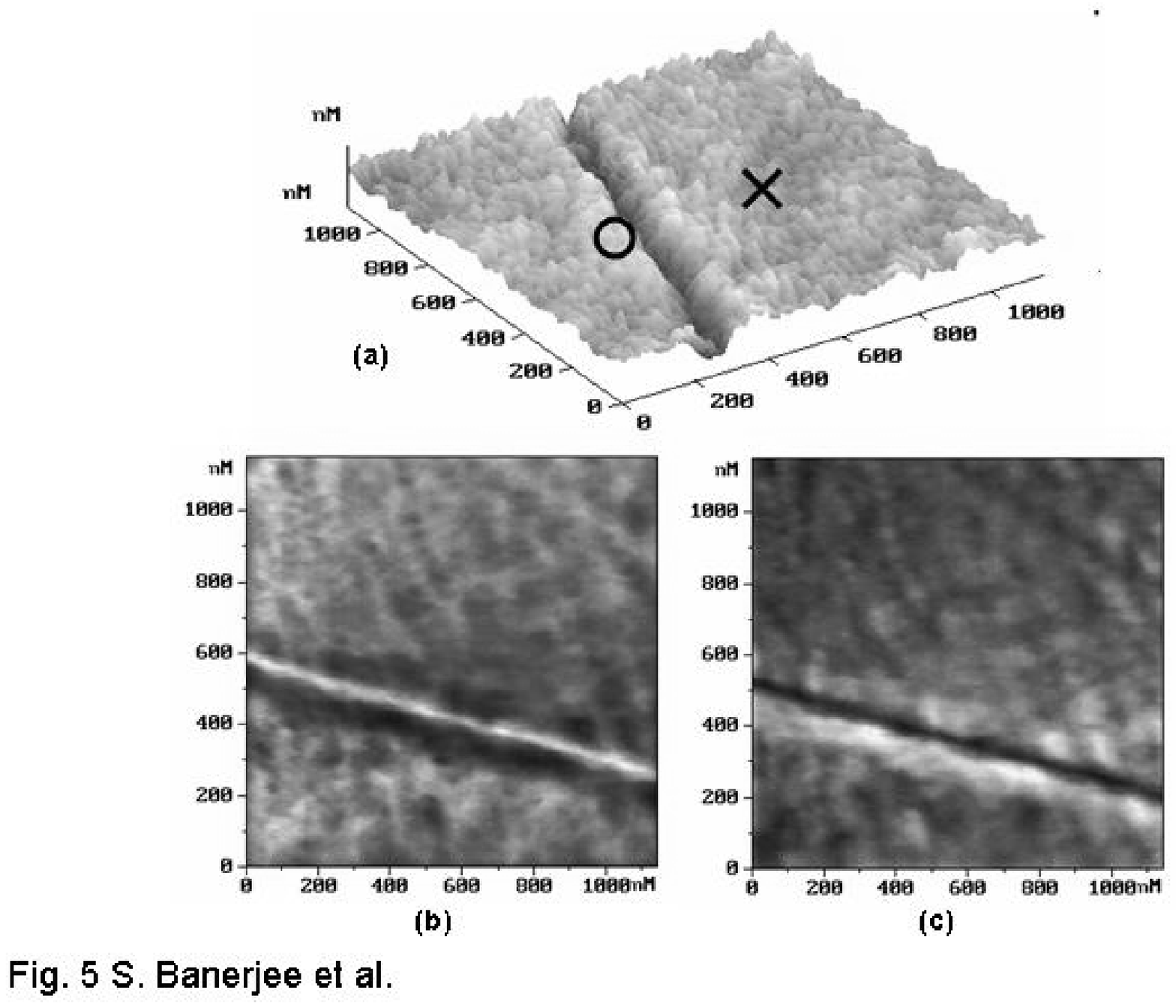}
\caption{(a) We show topography of the scratched film of gold (Au) deposited on silicon substrate. The feature we observe is of a deleberately carefully scratched region for the study of AFAM. (b) and (c) are the AFAM measurement carried out above and below the resonance curve respectively. Bright region in the AFAM image (b) corresponds to stiff region and dark region as soft regions. In image (c) the dark region corresponds to stiff region and the bright region corresponds to soft region. We observe contrast inversion in (b) and (c). Force vs. distance measurement were carried out at the regions marked by cross on the unscratched region and circle on the debris region shown in image (a).}
\end{figure}

\begin{figure}
\includegraphics*[angle=270,width=12cm]{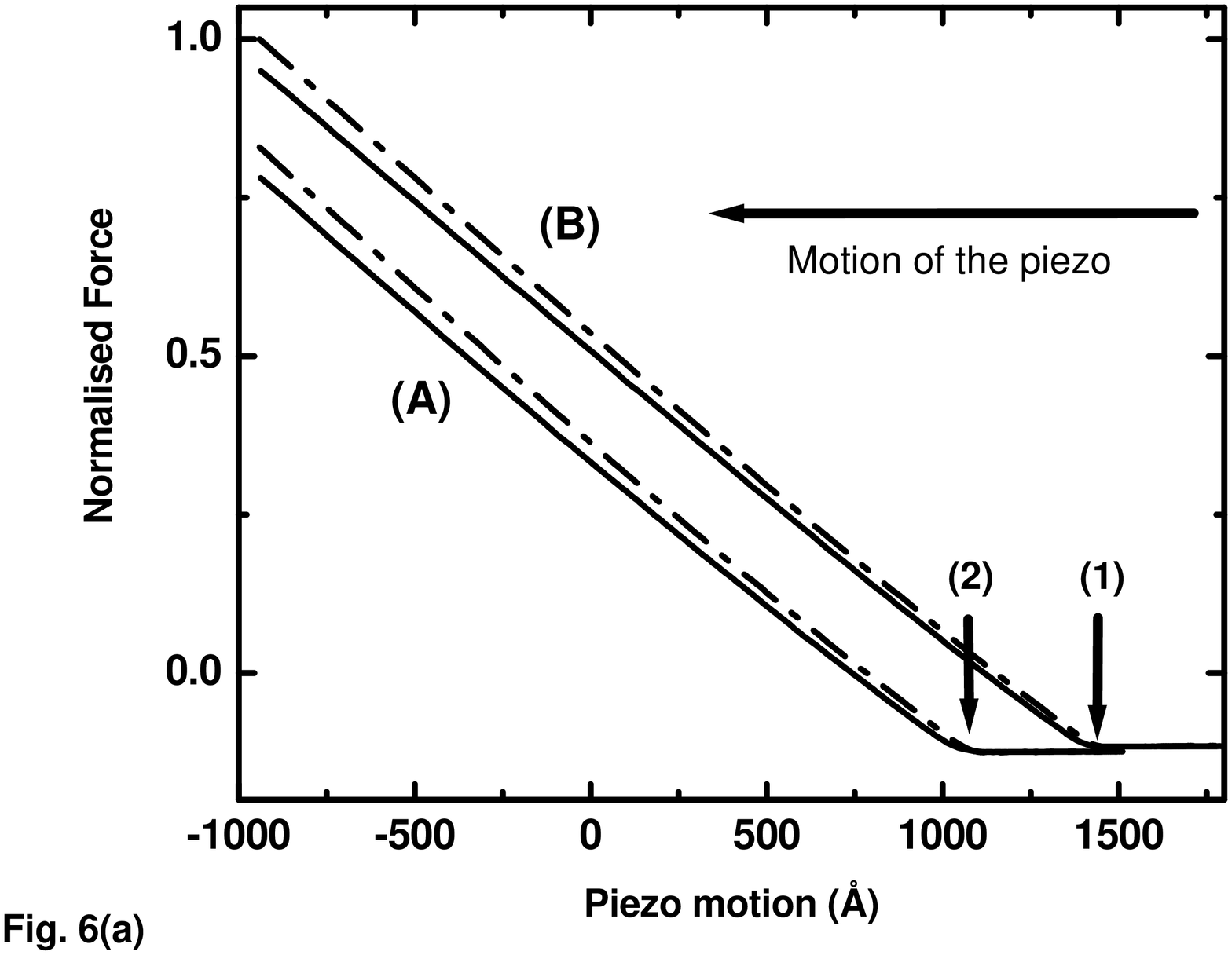}
\includegraphics*[angle=270,width=12cm]{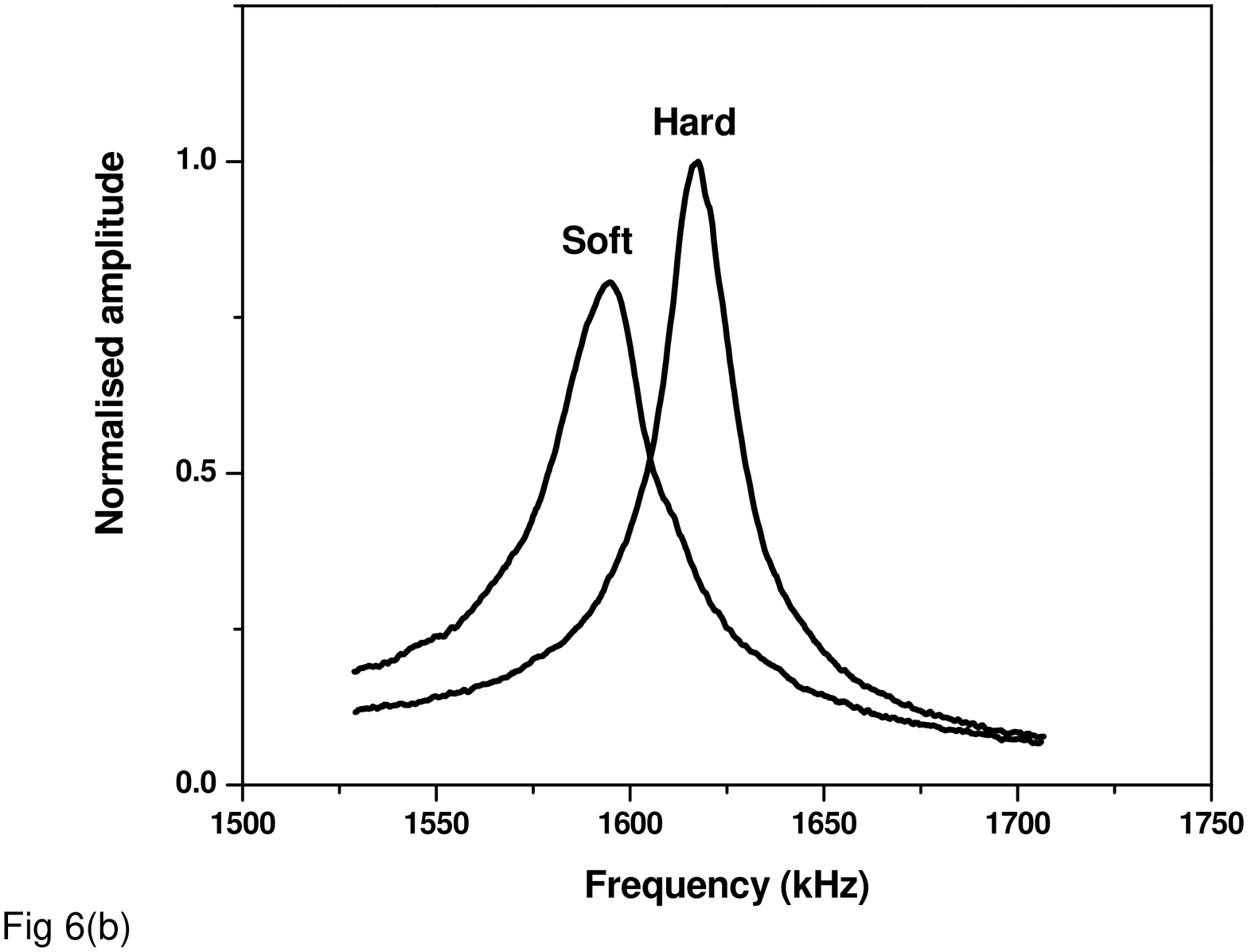}
\caption{(a) The Force vs distance curve on the unscratched region marked in Fig. 5(a) by cross and is shown here as a dot-dash line and on the peeled debris region marked by circle in Fig. 5(a) is shown as a solid line. The curve A and B are the measurement carried out with initial seperation of cantilever-sample having high and low value respectively (b) We show here the resonance curve taken on the unscratched (hard) region and on the debris (soft) region. We observe for the debris that the peak of the resonance curve is lower than on the unscratched region, indicating that the debris is softer than the unscratched region.}
\end{figure}

\end{document}